\documentclass[fleqn,usenatbib]{mnras}

\usepackage{newtxtext,newtxmath}

\usepackage[T1]{fontenc}
\usepackage{ae,aecompl}
\usepackage{float}

\usepackage{graphicx}	
\usepackage{amsmath}	
\usepackage{pdflscape}
\usepackage{booktabs}
\usepackage{multirow}

\usepackage{xcolor}
\usepackage[flushleft]{threeparttable}
\usepackage[normalem]{ulem}

\usepackage{tabularx}
\usepackage[normalem]{ulem}

\def\OIII{[O\,{\sc iii}]}

\title[A study of CLiF AGNs]{Coronal Line Forest AGN II: analysis of the spectral energy distribution}

\author[F. C. Cerqueira-Campos et al.]{F. C. Cerqueira-Campos$^{1}$\thanks{E-mail: fernando.campos@inpe.br},
A. Rodr\'iguez-Ardila$^{1,2}$\thanks{Visiting Astronomer at the Infrared Telescope Facility, which is operated by the University of Hawaii under contract NNH14CK55B with the National Aeronautics and Space Administration.},
S. Panda$^{2,4}$,
R. Riffel$^{3,6}$,
\newauthor{L. G. Dahmer-Hahn$^{5,2}$},
M. Marinello$^{2}$
\\
$^{1}$Divis\~ao de Astrof\'isica, Instituto Nacional de Pesquisas Espaciais, Avenida dos Astronautas 1758, S\~ao Jos\'e dos Campos, 12227-010, SP, Brazil\\
$^{2}$Laborat\'orio Nacional de Astrof\'isica - Rua dos Estados Unidos 154, Bairro das Na\c c\~oes. CEP 37504-364, Itajub\'a, MG, Brazil\\
$^{3}$Departamento de Astronomia, Instituto de F\'{\i}sica, Universidade Federal do Rio Grande do Sul, Av. Bento Gon\c calves, 9500. Porto Alegre, RS, Brazil\\
$^{4}$Center for Theoretical Physics, Polish Academy of Sciences, Al. Lotnik\'ow 32/46, 02-668 Warsaw, Poland\\
$^{5}$Shanghai Astronomical Observatory, Chinese Academy of Sciences, 80 Nandan road, Shanghai 200030, China\\
$^{6}$ Instituto de Astrof\'\i sica de Canarias, Calle V\'\i a L\'actea s/n, E-38205 La Laguna, Tenerife, Spain}

\date{Accepted XXX. Received YYY; in original form ZZZ}

\pubyear{2019}

\begin{document}
\label{firstpage}
\pagerange{\pageref{firstpage}--\pageref{lastpage}}
\maketitle

\begin{abstract}
Coronal-Line Forest Active Galactic Nuclei (CLiF AGN) are characterized by strong, high-ionization lines, which are in contrast to what is found in typical AGNs. Here, we carry out an infrared analysis aimed at understanding the spectral energy distribution of six sources from this group. In this work, the properties of the dusty torus for these objects are analyzed. To this purpose, we infer the physical and geometrical properties of the dust structure that surrounds the central region by fitting with models the spectral energy distribution (SED) of CLiF AGNs in the infrared. For this analysis, we compare the results of three models: {\sc clumpy}, {\sc skirtor} and {\sc cat3d-wind}. Using the Bayesian information criterion, {\sc skirtor} was found to have the most robust fit to the SEDs in five out of six galaxies. The remaining object was best fitted with {\sc clumpy}.  The results indicate that these objects are preferentially Type~I sources, supporting the detection of broad components in the permitted lines, likely associated with the BLR in the near-infrared (NIR) spectra. The best SED fitting indicates that the line of sight gives access to the view of the central source for these objects, but the amount of dusty clouds in the same direction is high, suggesting the hypothesis that they obscure the emission of the continuum produced by the central source and that the obscuration makes the coronal lines to not overlap with the continuum. 
 
\end{abstract}




\section{Introduction}\label{ch:cap1}

According to the unified model \citep{antonucci/1985}, the different spectral characteristics of active galactic nuclei (AGNs) are explained by the relative orientation between the axis of a dusty torus that surrounds the central source and the observer. The bulk of this structure is responsible for absorbing a significant fraction of the optical/ultraviolet continuum produced by the central source and re-emitting it in infrared wavelengths. Under this assumption, the diversity of spectral features observed in Type~I and Type~II AGNs is directly dependent on the line of sight between the central source and the observer. Observational evidence that the unresolved region of the AGNs is surrounded by an obscuring dusty structure abounds in the literature \citep{tristram/2007,kishimoto/2011,hoenig/2012,honig/2013,burtscher/2013,tristram/2014,honig/2014,lopez/2016,garcia/2019,leftley/2021,garcia/2021}. Moreover, the data suggest that the structure of this obscuring material in some cases is not only associated with a hot equatorial dust disk but also with polar dust clouds \citep{braatz/1993,cameron/1993,asmus/2019, honig/2019}. This polar dust is located on scales from tens to hundreds of parsecs from the central source \citep{asmus/2016}. The works of \cite{muller/2011} and \cite{riffel/2021} indicate that the coronal line region (CLR) is situated on this same scale.

The emission of coronal lines (CLs) in AGNs is mainly associated with the existence of highly energetic processes of the central engine. The study of the emitting region of these lines is important to trace the influence of the AGN in the host galaxy. 
CLs are not uncommon in AGN spectra, but not all AGNs display them.  \citet{riffel/2006}, in a NIR study of a sample of 47 AGNs, found that in $\sim$68\% of the objects, only one CL is identified (being they the emission lines of [\ion{S}{viii}], [\ion{S}{ix}], [\ion{Si}{vi}], [\ion{Si}{x}] and [\ion{Ca}{viii}]). Overall, in the sources where these lines are detected, they tend to be weak. In the optical regime, the typical values for the log ratio [\ion{O}{iii}] $\lambda$5007 / [\ion{Fe}{vii}] $\lambda$6087 is $-$0.24 $\pm$ 0.04 and $-$0.44 $\pm$ 0.06 for Seyfert 1 and Seyfert 2 respectively \citep{gelbord/2009}. It is important to notice that [\ion{Fe}{vii}] $\lambda$6087 is one of the most intense CLs in the optical spectrum.

\cite{rose/2015a}, \cite{rose/2015b} and \cite{glidden/2016} introduced in the literature a new class of AGNs dubbed Coronal Line Forest AGNs (CLiF AGNs). The optical spectrum of these objects is characterized by a plethora of unusually bright CLs. \cite{rose/2015a} claimed that in these particular sources, the CLs are produced in the inner wall of the dusty torus that surrounds the supermassive black hole (SMBH), with the assumption that at the limit of the radius of dust sublimation, a more refractive dust region is being intensely irradiated, undergoing ablation. According to \cite{krolik/1996}, under these conditions, thermal winds of highly ionized gas emerge and produce the observed coronal line forest. If this scenario is correct, in order to detect the coronal line region in CLiF AGNs, the relative orientation between the inner wall of the torus and the observer is restricted to a short range of intermediate angles, between that of Type~I and~II AGNs.

In \defcitealias{cerqueira/2021}{Paper~I}\citet[][hereafter Paper~I]{cerqueira/2021},  an extensive analysis of CLiF AGNs regarding the optical/near-infrared continuum emission, their spectral classification, extinction, gas kinematics and determination of their black hole masses, was carried out. The results obtained indicated the presence of broad components in the permitted emission lines in the NIR spectra, allowing to re-classify most CLiFs sources as Type~I AGNs. Through comparisons with line ratios and coronal lines luminosity with other non-CLiF AGNs, it was possible to verify that there are no clear distinctions between CLiF and non-CLiF AGNs. The gas kinematics in CLiFs suggested a compact NLR, of the order of a few arcsecs centered at the central source (<500 pc) (\citetalias{cerqueira/2021}). The black hole masses for this sample lie in the range between 10$^{7}$ - 10$^{8}$ M$_{\odot}$, supporting the hypothesis that these sources have hotter accretion disks \citep{panda/2018, prieto/2022}.  In the optical region, the luminosities found for CLs in CLiF AGNs are between 10$^{40}$-10$^{41}$ erg\,s$^{-1}$ (\citetalias{cerqueira/2021}). These values are in the upper range of luminosity reported by \citet{gelbord/2009} in a sample of 63 AGNs. Therefore, although not a separate class of AGNs, CLiFs tend to show extreme values in the properties studied. 

The spectral signature of the dusty torus is highly prominent in the infrared. Emission from hot dust from the AGN extends beyond 10 $\mu$m to $\sim 30 \mu$m \citep{rieke/1975}. However, observational evidence supports models that describe the 1 to 10 $\mu$m continuum as being predominantly or entirely due to hot dust emission, without major influences of the emission from the host galaxy (\citealt{edelson/1986,barvainis/1987,alonso/2003,martinez/2017}). Therefore, fitting the infrared spectral energy distribution (SED) is a good approach to infer the physical properties of the dusty structure around the central source. The analysis of \citet{rieke/1981} using models generated from large aperture spectra (approximately 8"), concluded that the emission in the NIR is dominated by thermal radiation produced by hot dust with a temperature between 1300 - 1500 K.  More recently, new models  available in the community better reproduce the observed SEDs (\citealt{fritz/2006,nenkova/2002,nenkova/2008b,honig/2010,siebenmorgen/2015,stalevski/2016,Honig/2017}). 

The torus models so far available in the literature can be divided into three classifications: smooth, clumpy and two-phase models. Initially, the first models were simpler and had a continuous or smooth distribution of dust in the torus, with a distribution of different radial and vertical density profiles (\citealt{pier/1992,granato/1994,efstathiou/1995,fritz/2006}). As dust would hardly survive in the medium with a unique continuous distribution, new models with dust organized into clouds were developed (\citealt{nenkova/2002,nenkova/2008b,van/2003,honig/2010, Honig/2017}). In addition, two-phase models, which use a scenario with the distribution of dust in both ways together, smooth and clumpy, were available in the literature (\citealt{stalevski/2012, siebenmorgen/2015,stalevski/2016}).

In Paper~I, we verified that CLiF AGNs do not display spectroscopic  properties that make them different from the distribution of non-CLiF AGNs. However, the physical mechanisms that produce such prominent CLs are still a matter of debate. Thus, in this second paper, the properties of the dusty torus in these objects are analyzed. To this purpose, we derive the physical and geometric properties of the dusty structure by fitting physically-motivated components to recover their SEDs in the infrared.   

This paper is organised as follows: We detail in Section \ref{dados} the sample and observations. In Section \ref{results}  we present the SED analysis using the {\sc clumpy}, {\sc skirtor} and {\sc cat3d-wind} models (\citealt{nenkova/2008b,stalevski/2016,Honig/2017}). In order to expand the discussion, we compare our results with those gathered for non-CLiF AGNs in Section \ref{compAnelise}. Main conclusions are drawn in Section \ref{conclusions}.

\section{Sample, observations and data reduction}\label{dados}

The sample chosen for this analysis consists of the six CLiF AGNs out of the seven already identified by \cite{rose/2015a} observed using NIR (7800 \AA\ $\leq$ $\lambda$ $\leq$ 25000 \AA) spectroscopy and mid- and far-infrared photometry. The objects identification are: ESO138-G001, SDSS\,J164126.91+432121.5, III\,Zw\,77,   MRK\,1388, SDSS\, 2MASX J113111.05+162739 and NGC\,424. The main characteristics of this sample can be found in \citetalias{cerqueira/2021}.

\subsection{NIR Spectroscopy}

The spectra of  SDSS~J124134.25+442639.2 and SDSS~J164126.91+432121.5 were obtained in queue mode with the 8.1\,m Gemini North telescope atop Mauna Kea using the Gemini Near-IR spectrograph \citep[GNIRS,][]{elias/2006} in the cross-dispersed mode. The spectra of III\,Zw~77 and Mrk\,1388 were obtained at the NASA 3\,m Infrared
Telescope Facility (IRTF) using the SpeX spectrograph \citep{rayner/2003} in the short
cross-dispersed mode (SXD, 0.7-2.4\,$\mu$m). Finally, he spectra of ESO~138-G001 and NGC\,424 were obtained using the ARCoIRIS spectrograph attached to the 4.1\,m Blanco Telescope \citep{schlawin/2014}. A more complete description of NIR spectroscopy for our sample is available on \citetalias{cerqueira/2021}.

\subsection{Mid and far infrared photometry}

We also employed photometric and spectroscopic measurements in the mid-infrared (MIR) and far-infrared (FIR) regions to construct the spectral energy distribution (SED) of the CLiF AGN sample to derive the torus properties. For NGC\,424 and ESO138-G001 these points were taken from spectra obtained by the public Spitzer Heritage Archive\footnote{\url{https://sha.ipac.caltech.edu/applications/Spitzer/SHA/}} data and reduced by SPICE\footnote{\url{https://irsa.ipac.caltech.edu/data/SPITZER/docs/dataanalysistools/tools/spice/}} (Spitzer IRS Custom Extraction). The instrument used for the observations was the IRS (InfraRed Spectrograph), equipped with a slit of 168 arcsecs in length by 10.7 arcsecs wide for the spectral range of 14-38\,$\mu$m. For the interval 7.4-14.5\,$\mu$m, Spitzer employs a slit of 57 $\times$ 3.7\,arcsecs.

For ESO\,138-G001, to minimize the effects of the host galaxy contribution, we adopted a similar method as that used by \cite{lira/2013} and \cite{audibert/2016}. It consists of the subtraction of the host galaxy contribution from the observed spectrum of templates developed by \cite{smith/2007}, where the MIR light is dominated by the contribution of star-forming regions. In this way, it is possible to obtain a more reliable representation of the spectrum emitted by the AGN. In addition, we removed any potential contribution of star formation using  templates and Gaussian representations of other forbidden emission lines. 
To this purpose we used the {\sc pahfit} tool developed by \cite{smith/2007} made for the low-resolution IRS Spitzer spectra. This software models the observed emission as the sum of the starlight continuum, thermal dust continuum, pure rotational lines of H$_{2}$, fine structure lines and dust emission features. In our case, as we are interested in isolating the AGN continuum, we decided to subtract only the emission lines from the H$_{2}$ and the molecular features of PAH (polycyclic aromatic hydrocarbons) emission, stellar contribution and the fine structure lines.

\begin{table}
\centering
  \caption{Photometric points in the MIR for the galaxy sample extracted from WISE and Akari observations.}
\resizebox{!}{3.2cm}{\begin{tabular}{llll}
\hline
Galaxy & $\lambda$ ($\mu$m) & Flux (mJy) & Origin:Filter \\ \hline
\textbf{SDSS J164126.90+432121.5} &22.09 & 48.2$\pm$1.6 & WISE:W4 \\
                                  &11.56 & 14.0$\pm$0.3 & WISE:W3 \\
                                  &4.60 & 3.7$\pm$0.1 & WISE:W2 \\
                                  &3.35 & 2.1$\pm$0.1 & WISE:W1 \\ \hline
\textbf{III Zw 77}& 22.09 & 114.0$\pm$3.0 & WISE:W4 \\
                  &11.56 & 49.8$\pm$0.7 & WISE:W3 \\
                  &4.60 & 17.4$\pm$0.3 & WISE:W2 \\
                  &3.35 & 12.3$\pm$0.2 & WISE:W1 \\ \hline
\textbf{MRK 1388} & 22.09 & 228.0$\pm$4.0 & WISE:W4 \\
                  & 18.39 & 122.0$\pm$52.0 & AKARI:L18W \\
                  &11.56 & 89.8$\pm$1.2 & WISE:W3 \\
                  &8.61 & 55.8$\pm$9.3 & AKARI:S9W \\
                  &4.60 & 22.8$\pm$0.4 & WISE:W2 \\
                  &3.35 & 14.5$\pm$0.3 & WISE:W1 \\ \hline
\textbf{2MASX J113111.05+162739} &22.09 & 40.5$\pm$1.8 & WISE:W4 \\
                                 &11.56 & 19.4$\pm$0.3 & WISE:W3 \\
                                 & 4.60 & 6.1$\pm$0.1 & WISE:W2 \\
                                 & 3.35 & 3.3$\pm$0.1 & WISE:W1 \\ \hline
\end{tabular}}
\label{tab: phot}%
\end{table}

\begin{figure}

  \begin{center}
  \includegraphics[width=\columnwidth]{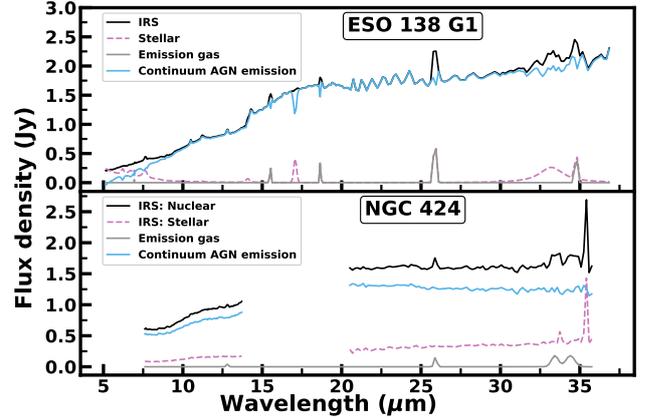}
    
  \caption{Spitzer spectra for ESO 138 G1 and NGC 424. The solid black line represents the nuclear spectrum plus the stellar component, the dashed magenta line represents the stellar component, the solid grey line represents the gas emission, and the solid blue line represents the central source continuum emission.}\label{fig: spitz}
    
  \end{center}
    
\end{figure}

Most of the emission of PAH lies in the 5-15~$\mu$m interval,  where the stellar contribution is most prominent. Unfortunately for longer wavelengths the separation of stellar emission and AGN continuum is not possible. Thus, this method may overestimate the nuclear emission for  $\lambda$ > 20 $\mu$m. The upper panel of Figure~\ref{fig: spitz} shows the observed spectrum, the model with the emission PAH and H$_2$ features and the residual after subtraction of that component.

For NGC\,424 we followed a different approach. Because of the large slit width of Spitzer, the spectrum covers not only the AGN but also the host galaxy. For sources with strong nuclear emission such as NGC 424, it was possible to accurately separate both components thanks to the method employed during the observations. The galaxy was mapped by a set of slits, using one to observe the central region and the host galaxy and another that does not pass through the centre. A more reliable flux density of the central region can be obtained by subtracting the emission with only the galaxy contribution from that with the nuclear and host galaxy components. In addition, we also applied the templates of \cite{smith/2007} to remove the pure rotational lines of H$_{2}$ and fine structure lines.
It is possible to verify in the bottom panel of Figure \ref{fig: spitz} the removal of the NGC 424 host galaxy contribution and nebular emission, resulting in a pure continuum AGN emission.

For  2MASX~J113111.05+162739, SDSS J164126.91+432121.5, III\,Zw\,77 and MKR 1388, photometric points were extracted from the VizieR catalog database of the space telescopes WISE and AKARI\footnote{\label{myfootnote} This research has made use of the VizieR catalogue access tool, CDS, Strasbourg, France (DOI: 10.26093/cds/vizier). The original description of the VizieR service was published in \citet{vizier-catalogue_2000}.}. The source of each observation and the fluxes are shown in table \ref{tab: phot}. 




\section{SED Fitting}\label{results}

\subsection{Set of models employed}\label{seds}
 
During the last two decades, several models have been developed to describe the SED of AGNs in the infrared, including the signatures that the dusty torus imprints on that distribution. In this work, we used three of these models and compare their results to identify the most likely parameters that best describe the geometry of the dusty torus for each object in the sample.

The first of them is the {\sc clumpy} model (\citealt{nenkova/2008a}, \citealt{nenkova/2008b}), which examines the infrared emission of the torus as a function of the inclination angle with respect to the observer. It assumes that the shape of the dusty torus is not homogeneous but consists of an ensemble of discrete dust clouds. In this scenario, we expect the infrared emission at shorter wavelengths to decrease progressively more than the emission at longer wavelengths as we reach closer to the edge-on orientation. This is due to a combination of an increasing number of clouds intercepted by the line of sight and a higher absorption at the shorter wavelengths relative to the longer wavelengths.

The {\sc clumpy} models have six free parameters: the number of clouds in the direction of the line of sight ($N_{0}$); the ratio between the outer and inner radius of the torus ($Y$); the power-law index describing the radial density of clouds of the torus ($q$); the angle between the center of the system and the outer wall of the torus ($\sigma$); the optical depth ($\tau _{\nu }$) and the viewing angle of the torus relative to the observer ($i$). The latter parameter is the most relevant one to our work.

The second model is {\sc skirtor} \citep{stalevski/2016}. It considers the torus to be in a two-phase composition, i. e., the torus geometry consists of a set of high-density dusty clumps immersed in a smooth dusty medium of relatively low density.

The {\sc skirtor} models have six free parameters. Like {\sc clumpy}, it has the parameters $i$, $Y$ and $\sigma$. In addition, it uses the parameters that represent the edge-on optical depth at 9.7 microns ($\tau _{9.7\mu m}$); a power-law index that sets the dust density gradient for the radial ($p$) and polar ($q$) angle. The fraction of total dust mass inside clumps is set to 97$\%$ with the remaining 3$\%$ for the smooth interclump medium. It does not consider the presence of polar dust clouds.

The third model is {\sc cat3d-wind} (\citealt{Honig/2017}). The main feature that differentiates it from {\sc clumpy} is that the former adds a polar outflow, modelled as a hollow cone, which implies a radial distribution of dusty clouds along a polar wind. 

The parameters employed to characterize {\sc cat3d-wind} are: The index of the power-law ($q$) that describes the radial dust cloud distribution; the half-opening angle ($\sigma$); the number of clouds along an equatorial line of sight ($N_{0}$); the index of the dust cloud distribution power-law along the polar wind ($a_{w}$); the half-opening angle of the wind ($\theta_{w}$); the angular width of the hollow wind cone ($\sigma_{\theta}$); the ratio of dust clouds along the line of sight of the wind compared to the one in the disk plane ($f_{wd}$); the outer radius of the torus ($R_{out}$); the optical depth of the individual clouds ($\tau$), and the viewing angle of the torus relative to the observer ($i$). 

In order to build the SED for each object of the sample in the MIR and FIR, Spitzer spectra, WISE and AKARI photometric points were employed. For the NIR and optical regions, integrated spectra were used. The spectral data source information is described in section \ref{dados}. Additional information on the spectroscopy in the NIR  is provided in \citetalias{cerqueira/2021}. The emission lines of the spectra were removed for the fit, so that we only employed the continuum emission.

Since difefrent wavelengt ranges are dominated by different components, we first fitted the optical range in order to propagated this component to the remaining regions. This choice was done based on the fact that the optical continuum is expected to be simpler, being composed of a combination of stellar populations plus a featureless continuum (FC) produced by the AGN. In order to determine the different contributions to the observed continuum in the optical/NIR for our sample (i.e. stellar population and AGN), we have employed the {\sc starlight} code \citep{CF+04, CF+05}. Basically, the code fits an empirical stellar library to the observed spectrum by minimizing the $\chi^2$, taking into account factors like stellar kinematics and dust reddening. 

We fed {\sc starlight} with a modified version of the \citet{BC03} library of simple stellar populations, with \citet{Chabrier+03} IMF and Padova \citep{Girardi+00} isochrones. We have limited the library to the 25 most representative ages\footnote{0.00100, 0.00316, 0.00501, 0.00661, 0.00871, 0.0100, 0.0144, 0.0251, 0.0400, 0.0550, 0.101, 0.161, 0.286, 0.509, 0.904, 1.27, 1.43, 2.50, 4.25, 6.25, 7.50, 10.0, 13.0, 15.0 and 18.0~Gyr} and 6 most representative metallicities\footnote{Z=0.00010, 0.00040, 0.00400, 0.00800, 0.02000 and 0.05000}. We chose this library because it has a very wide wavelength coverage, encompassing most wavelengths analysed in this paper. In order to represent the contribution from the AGN, we added a power law following the expression f$_\lambda \propto \lambda^{-0.5}$. An example of the application of the code and method to the optical and NIR can be found in \citet{luisgdh+19a}. We used the stellar component added to the models mentioned above to build the SED. We then created a NIR+MIR stellar and AGN continuum by fixing the stellar populations and featureless continuum fractions derived from the optical, and propagating this continuum to the remaining wavelength ranges.

After determining the best stellar template and the fraction that it contributes in each galaxy, we employed the {\sc python} package {\sc lmfit} to fit the SED (\citealt{newville/2015}). {\sc lmfit} provides a high-level interface to non-linear optimization and curve-fitting problems. Using this routine, it was possible to identify the best model with respect to the points obtained through observations. 

We also employed, for comparison purposes, photometric points for NGC\,4151 and NGC\,1068 as representatives of Type~I and~II AGN, respectively. The VizieR photometry tool$^{\ref{myfootnote}}$ was used to gather the data. For NGC\,1068 in the optical range, a spectrum of the inner 500 × 500 pc obtained from the Multi Unit Spectroscopic Explorer (MUSE) processed archival data was used and the stellar continuum was estimated using the {\sc starlight}. The contribution of the stellar population in NGC\,4151 is estimated at 15$\%$ in the NIR spectrum \citep{riffel/2009}, therefore the same stellar population template of NGC\,1068 was used but normalized to represent 15$\%$ of the flux in the NIR for this galaxy SED.

\subsection{Criterion to choose the best fit}\label{fitandoSed}

\begin{figure}
    \centering
    \includegraphics[width=\columnwidth,height=6cm]{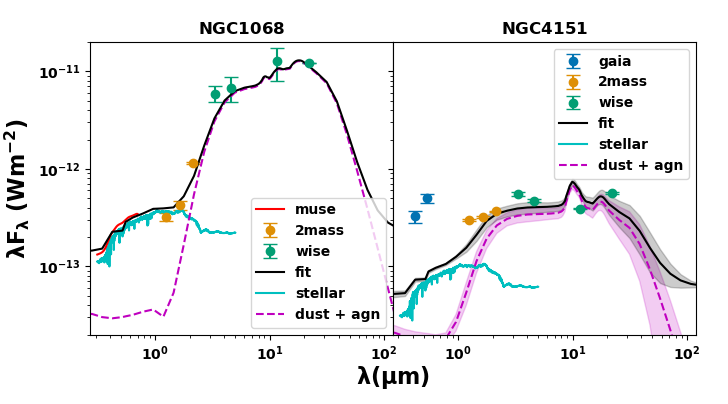}
    \caption{Best fit from {\sc clumpy} models to the archetypal Seyfert\,1 and\,2 galaxies NGC\,4151 (right panel) and NGC\,1068 (left panel), respectively. The black line represents the fitted SED. The points represent the observational data from Gaia (blue), 2MASS (yellow) and Wise (green), the red line represents the MUSE data. The shaded area represents the dispersion in the fitted SED.}
    \label{fig:sedst1e2clumpy}
\end{figure}

\begin{figure}
\centering
    \includegraphics[width=\columnwidth,height=6cm]{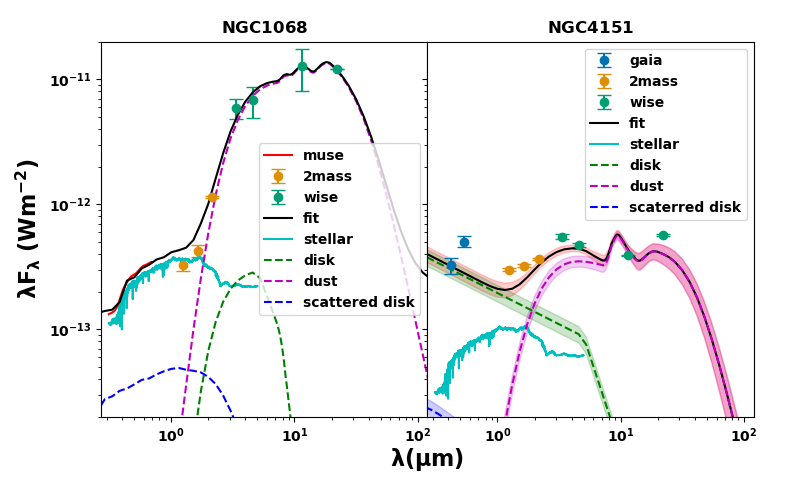}
    \caption{Same as Figure \ref{fig:sedst1e2clumpy} but for the {\sc skirtor} model.}
	\label{fig:sedst1e2skir}
\end{figure}

\begin{figure}
\centering
    \includegraphics[width=\columnwidth,height=6cm]{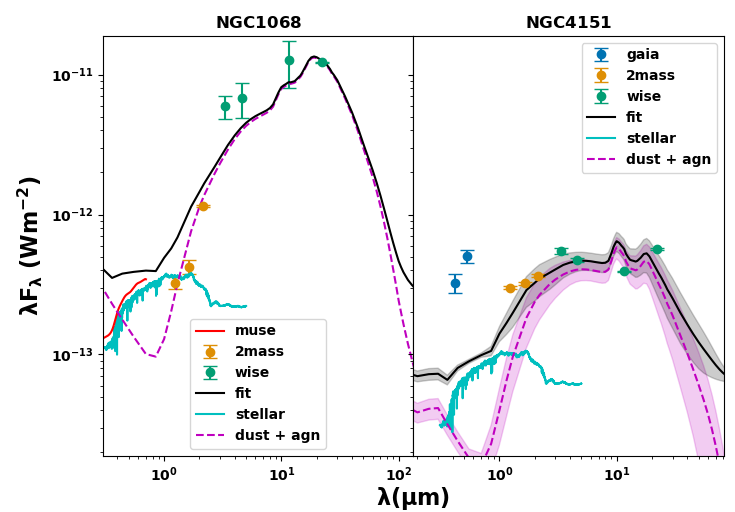}
    \caption{Same as Figure \ref{fig:sedst1e2clumpy} but for the {\sc cat3d-wind} model.}
	\label{fig:sedst1e2cat}
\end{figure}

\begin{figure}
    \begin{center}
    \includegraphics[width=\columnwidth]{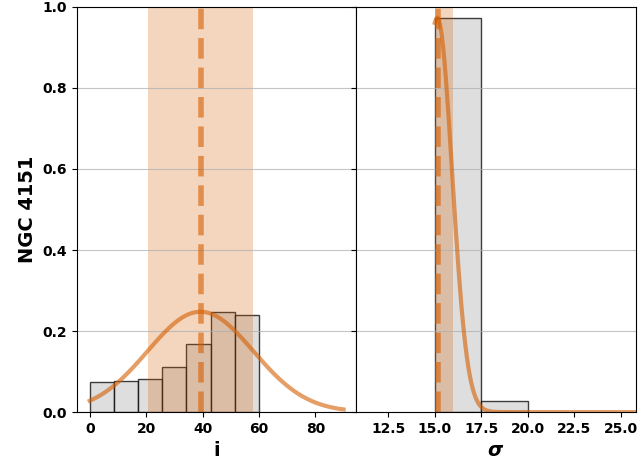}
    \caption{Histogram of the best value for the $i$ and $\sigma$ parameters for {\sc clumpy} in the archetypal Seyfert\,1  NGC\,4151. The orange-dashed line represents the most probable value. The orange solid line is the distribution function, and the vertical orange area the dispersion in the respective value.}\label{fig:histclumpy}
    \end{center}
\end{figure}

\begin{figure}
    \begin{center}
    \includegraphics[width=\columnwidth]{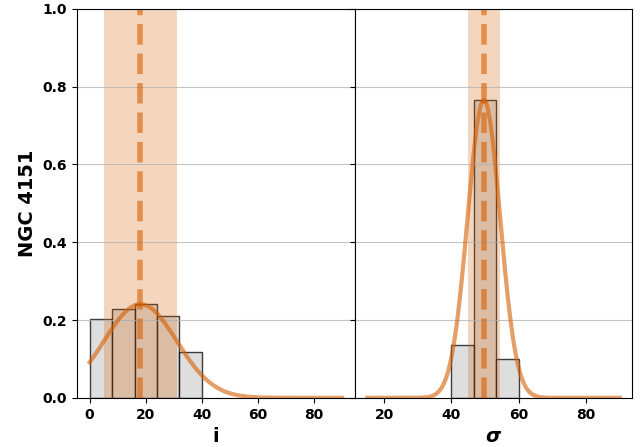}
    \caption{Same as Figure \ref{fig:histclumpy} but for {\sc skirtor}}\label{fig:histskir}
    \end{center}
\end{figure}

\begin{figure}
    \begin{center}
    \includegraphics[width=\columnwidth]{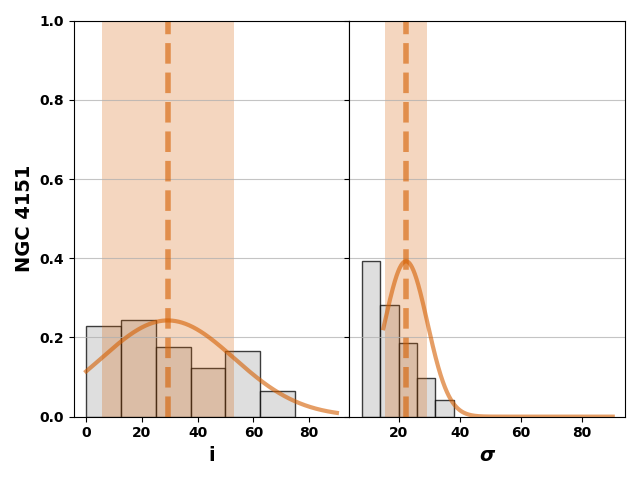}
    \caption{Same as Figure \ref{fig:histclumpy} but for {\sc cat3d-wind}}\label{fig:histcat3d}
    \end{center}
\end{figure}

\begin{figure*}
    \includegraphics[width=16cm,height=10cm]{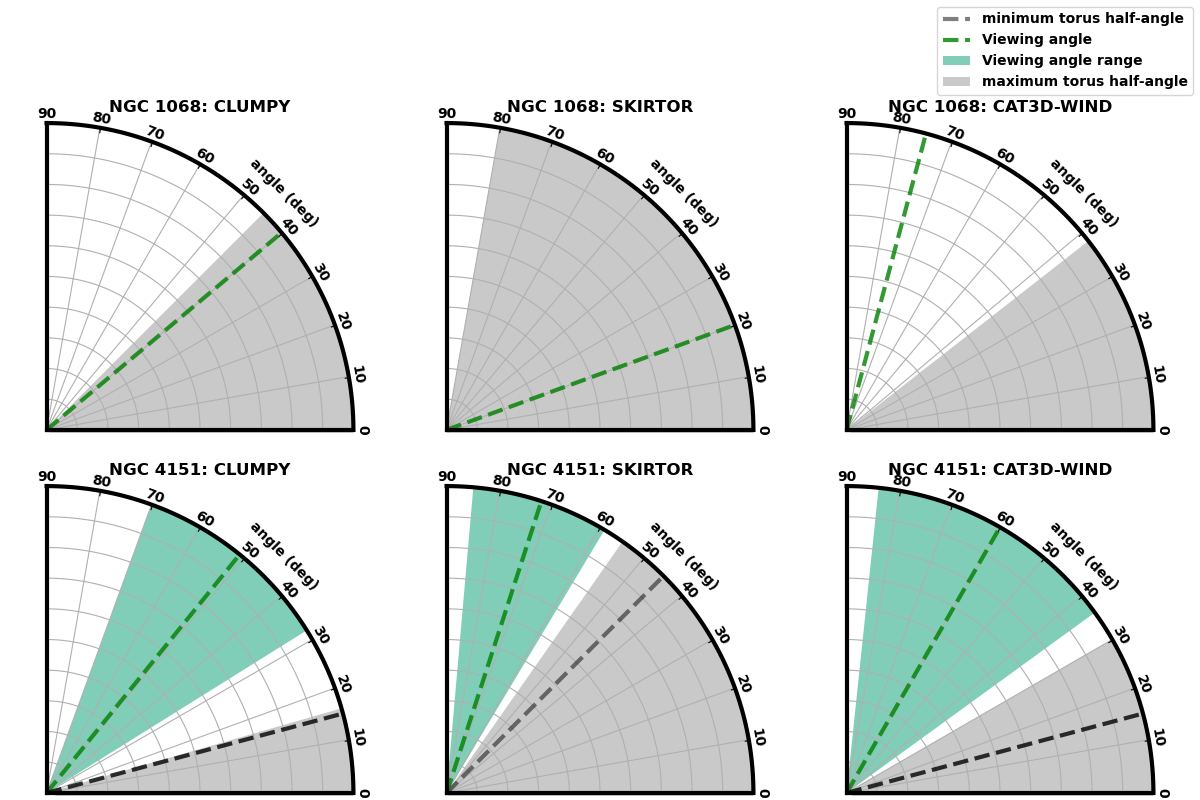}
    \caption{The representation of the $\sigma$ and $i$ parameters for the {\sc skirtor} and {\sc clumpy} models to the galaxies NGC\,1068 and NGC\,4151. The dashed green line represents the best-fit value for the viewing angle and the green area marks the error in the estimation of the viewing angle. The grey area marks the maximum range of the half-opening angle of the torus and the dashed grey line represents the minimum value of the half-opening angle of the torus. In this representation the angle is 0 deg. represents the torus equatorial plane}
	\label{fig:polart1e2}
	\centering
\end{figure*}
 
The {\sc clumpy}, {\sc skirtor} and {\sc cat3d-wind} models consist of 1.247.400, 19.200 and 124.740 SED templates, respectively. They are constructed through different combinations of the parameters described in section~\ref{seds}. 

To build the SED of the objects in our sample, we used the method described in Section~\ref{seds}. Moreover, we developed an algorithm in {\sc python} to fit the SEDs using the {\sc lmfit} package (\citealt{newville/2015}).  It uses a normalization factor for each template in order to obtain the best fit for the observational data. For each fit, the code returns a Bayesian information criterion (BIC) list, with the best fits having the lowest BIC. However, because the solution from the models is intrinsically degenerate, it is necessary to take into account not only the smallest BIC to obtain more realistic results, but also to choose a sample of best fits that do not drastically oppose the best result.

Thus, the final result for each source is the average of the best models that fit the observational data. In this process, we followed the procedure described by \cite{kass/1995}. Since the best model is the one that provides the minimum BIC, which we will denote by BIC*, the variation $\Delta$BIC = BIC$_{m}$ - BIC* is an indicator of opposition to the best model, $m$ being the number of models. With $\Delta$BIC $<$ 6, the model with the best result presents a positive but not a strong opposition. Thus, we selected the set of models that confronts the smallest model with a variation of less than 6 to compose the average SED model.

\subsection{Test case with NGC 4151 and NGC 1068}\label{test}

In order to test the fitting procedure  with {\sc clumpy},  {\sc cat3d-wind} and {\sc skirtor}, we first fit the SED of the archetypal Seyfert\,1 and\,2 galaxies NGC\,4151 and NGC\,1068, respectively. To compose the SED, photometric points in the optical, NIR, and MIR  were extracted from Gaia,  2MASS, and  WISE, respectively, through the VizieR photometry tool. 

The best SED fitting obtained for NGC\,4151 and NGC\,1068 are illustrated in Figure \ref{fig:sedst1e2clumpy} for {\sc clumpy}, Figure \ref{fig:sedst1e2skir} for {\sc skirtor} and in Figure \ref{fig:sedst1e2cat} for {\sc cat3d-wind}. In order to evaluate the consistency of the results in relation to the spectral type, the parameters $\sigma$ and $i$ were analyzed. For NGC\,4151, as only photometric points were used, the problem becomes more degenerate and a group of templates fit in the same model present some opposition with respect to $\Delta$BIC.

In order to find the most probable value and the uncertainty for each parameter, it was necessary to identify the best function that describes the data sample distribution of each variable. To this purpose, the python package {\sc fitter} was used \citep{Cokelaer/2021}. It allows the determination of the most probable distribution and the best value of each parameter. We used the lower BIC as a selection criterion among a range of distribution functions (Norm, Rayleigh, Cauchy, Power Norm, $\chi ^{2}$, Power Lognormal, t-distribution, F-distribution, Weibull Max). 

In figures \ref{fig:histclumpy}, \ref{fig:histskir} and \ref{fig:histcat3d}, we plot the distribution of values of the inclination angle and sigma found using {\sc clumpy}, {\sc skirtor} and {\sc cat3d-wind}, respectively, for NGC\,4151. In each panel, we show the histogram of the distribution of the parameter values of the models that fulfilled the $\Delta$BIC selection criterion. The solid orange line represents the function that describes the best distribution, the vertical orange dashed line represents the most probable value and the orange shaded area is the uncertainty range with 1$\sigma$ confidence.

For NGC\,4151 the BIC value of the best-fitted template was -544, -530 and -527 for {\sc skirtor}, {\sc clumpy}, and {\sc cat3d-wind}, respectively. It is not possible to claim that the results are comparable. For the parameter $i$ the values obtained are 18 $\pm$ 13 using {\sc skirtor}, 39 $\pm$ 19 for {\sc clumpy} and 30$\pm$24 for {\sc cat3d-wind}. The {\sc skirtor} value for the viewing angle is compatible, within uncertainties, to 9$_{-9}^{+18}$ deg, reported by \cite{nandra/1997} using X-ray observations from ASCA.

For NGC\,1068 the BIC value of the best-fit template was -120624, -121683 and -114423 for {\sc skirtor}, {\sc clumpy} and {\sc cat3d-wind}, respectively. For the inclination angle $i$ we found 70$\pm$10 deg for {\sc skirtor} and 50$\pm$10 deg for {\sc clumpy}. The latter value, despite not having the lowest BIC, is in agreement within uncertainties, with those  derived in the literature. \citet{garcia/2016}, for example, inferred an inclination of 66$_{-4}^{+9}$ deg. \citet{lopez/2018} found a value of 75$_{-4}^{+8}$ deg and \citet{GravityCollaboration/2020} derived a value of 70 $\pm$ 5 deg. For the $i$ identified using {\sc cat3d-wind}, the model has the worst fit among the three models as the value of 15$\pm$15 predicted is not consistent with the Type~II nature of that source.

It is possible to verify that even employing photometric points from the optical to the MIR spectral windows, the results obtained with the {\sc skirtor} models are compatible with those in the literature. The outcome of {\sc cat3d-wind} is the least plausible among the three models, not representing the optical region very well in the two test objects. Moreover, regarding the spectral type, the models {\sc skirtor} and {\sc clumpy} present consistent results, except for the fit with {\sc clumpy} for the NGC\,4151, which fails at reproducing  the optical part of the SED. Figure~\ref{fig:polart1e2} displays a planar diagram with the parameters $i$ and $\sigma$ found using the three SED models. For NGC\,1068, both {\sc skirtor} and {\sc clumpy} suggest that we are looking at this source through the torus. For NGC\,4151, {\sc clumpy}, {\sc skirtor} and {\sc cat3d-wind} suggest a viewing angle such that the observer is facing the central source.

\subsection{SED fitting of CLiF AGNs}\label{fitcritSed}

After testing the consistency of our approach with two of the most $bona fide$ AGNs in the literature, we proceed to the fitting of the CLiF sample. Among the three torus models (i.e. {\sc clumpy}, {\sc skirtor}, and {\sc cat3d-wind}), to select the one that best describes each object in our sample, the lowest BIC value was used as a criterion. However, if there was a $\Delta$BIC $<$ 6 in the comparison of the BIC results, more than one model will be considered. 

In the fitting process, the contribution of   the stellar population templates generated by {\sc starlight} in the optical region (i.e, at 5400~\AA) was set to the values predicted by the stellar synthesis: 75$\%$, 97$\%$, 56$\%$, 75$\%$, 95$\%$, and 67$\%$ for ESO\,138\,G1, SDSS\,J164+43, III\,ZW\,77, MRK\,1388, 2MASX\,J113+16 and NGC\,424, respectively. Overall, it ranges from 56$\%$ to 97$\%$.

Table~\ref{tab:bics} shows the results of the lowest BIC for each model. When comparing among the three Torus models, there was no solution with $\Delta$BIC $<$ 6. Five objects had the best result with {\sc skirtor}, one with {\sc clumpy} and none with {\sc cat3d-wind}, suggesting that the polar wind component in these sources is not a prominent contributor to the overall SED. For this reason, the latter was not considered in the analysis that follows. The templates that fit with the lowest BIC value, compared to the second-best fit of the same model, also had a $\Delta$BIC value greater than six. Therefore, only the best-fitted template for each model was considered in the final results. For this reason, the error bars were estimated taking into account the binning of the parameters in the template models.

\begin{table}
\centering
 \caption{The lowest BIC results for each of the three models in the SED fitting. Values in blue are the best results while those in red the worst.}
\begin{tabular}{llll}
\hline
\multicolumn{1}{c}{Galaxy} & \multicolumn{1}{c}{{\sc clumpy}}        & \multicolumn{1}{c}{{\sc skirtor}}       & \multicolumn{1}{c}{{\sc cat3d-wind}}    \\ \hline
ESO 138 G1                          & {\color[HTML]{3531FF} \textbf{-792636}}   & -779617                                    & {\color[HTML]{333333} -781789}                \\
SDSS J164+43                        & {\color[HTML]{CB0000} \textbf{-656610}}   & {\color[HTML]{3531FF} \textbf{-690268}}    & -685362                                       \\
III ZW 77                           & -671932                                   & {\color[HTML]{3531FF} \textbf{-672257}}    & {\color[HTML]{CB0000} \textbf{-653336}}       \\
MRK 1388                            & {\color[HTML]{CB0000} \textbf{-739757}}   & {\color[HTML]{3531FF} \textbf{-761091}}    & {\color[HTML]{333333} -744685}                \\
2MASX J113+16                       & {\color[HTML]{CB0000} \textbf{-660216}}   & {\color[HTML]{3531FF} \textbf{-697847}}    & -688366                                       \\
NGC 424                             & {\color[HTML]{CB0000} \textbf{-688501}}   & {\color[HTML]{3531FF} \textbf{-733837}}    & -731986                                       \\ \hline
\end{tabular}
\label{tab:bics}
\end{table}

\begin{figure*}
    \includegraphics[width=16cm]{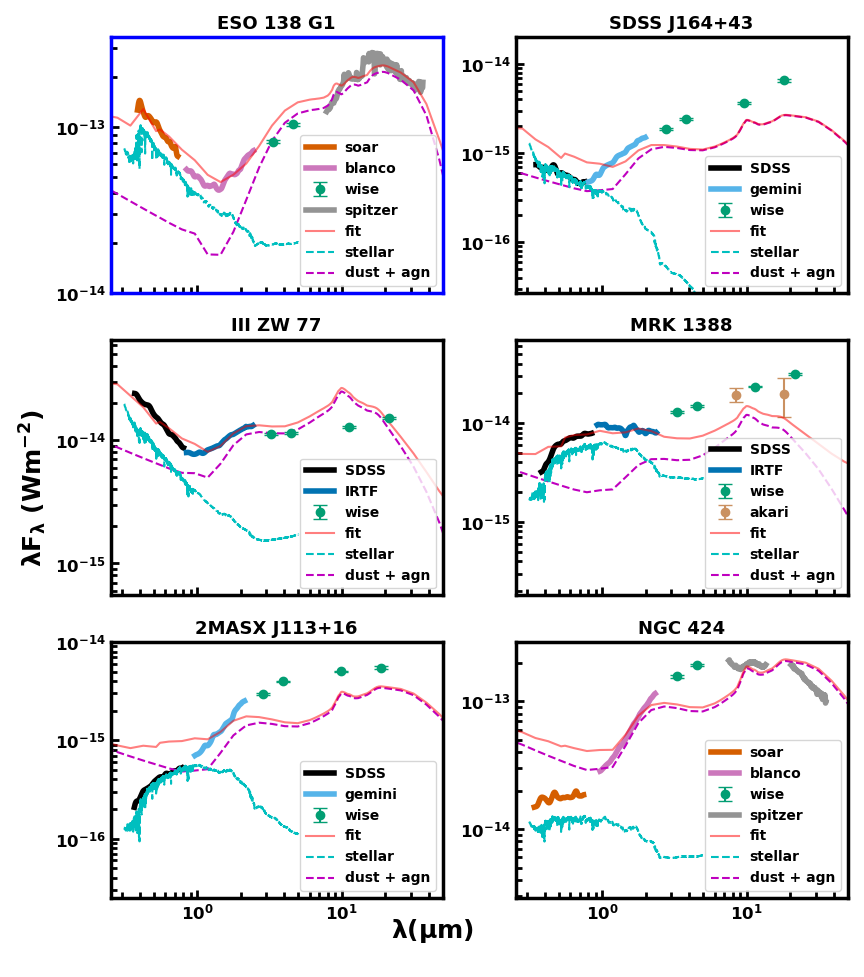}
    \caption{Best fit to CLiF AGN SEDs to {\sc clumpy} models. The cyan line represents the stellar component, the dashed magenta line represents the dust plus the central source emission and the red solid line is the best fit. The panel in blue represents the best fit among all other models. The x-axis represents the rest wavelength.}
	\label{fig:sed1}
	\centering
\end{figure*}

\begin{figure*}
    \includegraphics[width=16cm]{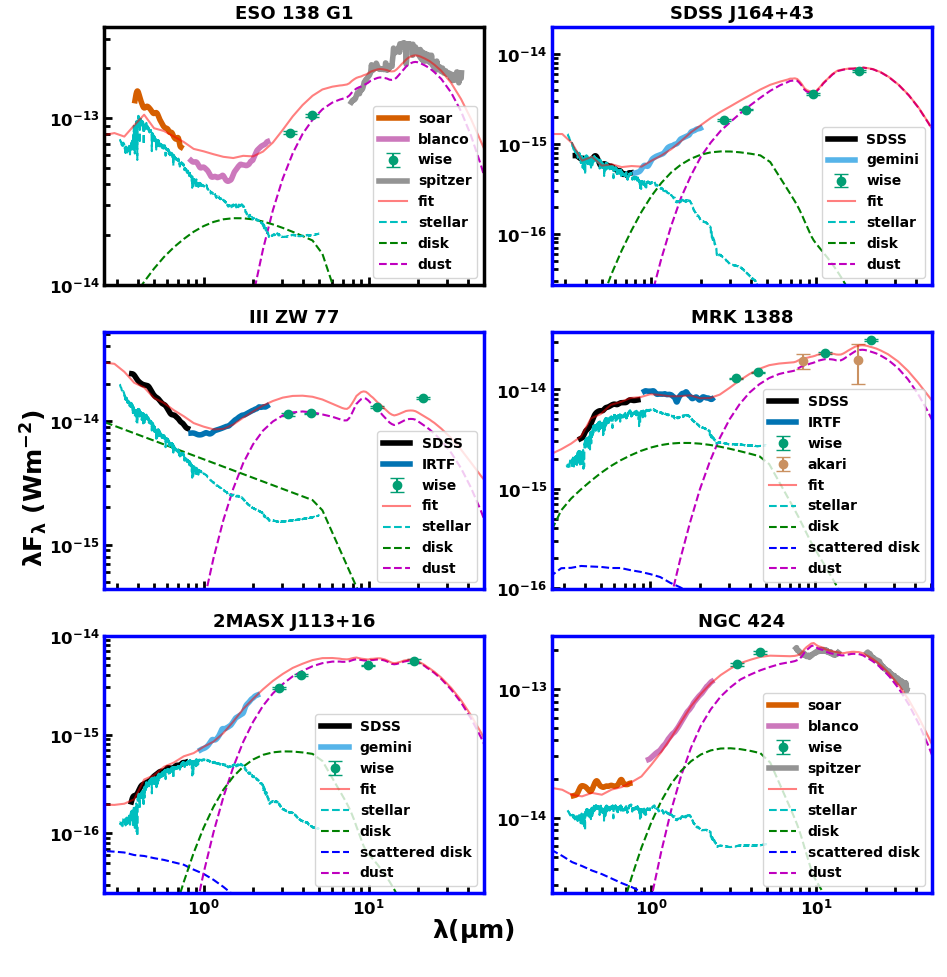}
    \caption{Best fit to CLiF AGN SEDs to {\sc skirtor} models. The cyan line represents the stellar component, the dashed magenta line represents the dust, the green dashed line represents the central source emission, the dashed blue line represents the scattered disk emission and the red solid line is the best fit. The panel in blue represents the best fit among all other models. The x-axis represents the rest wavelength.}
	\label{fig:sed2}
	\centering
\end{figure*}

\begin{figure*}
    \includegraphics[width=16cm]{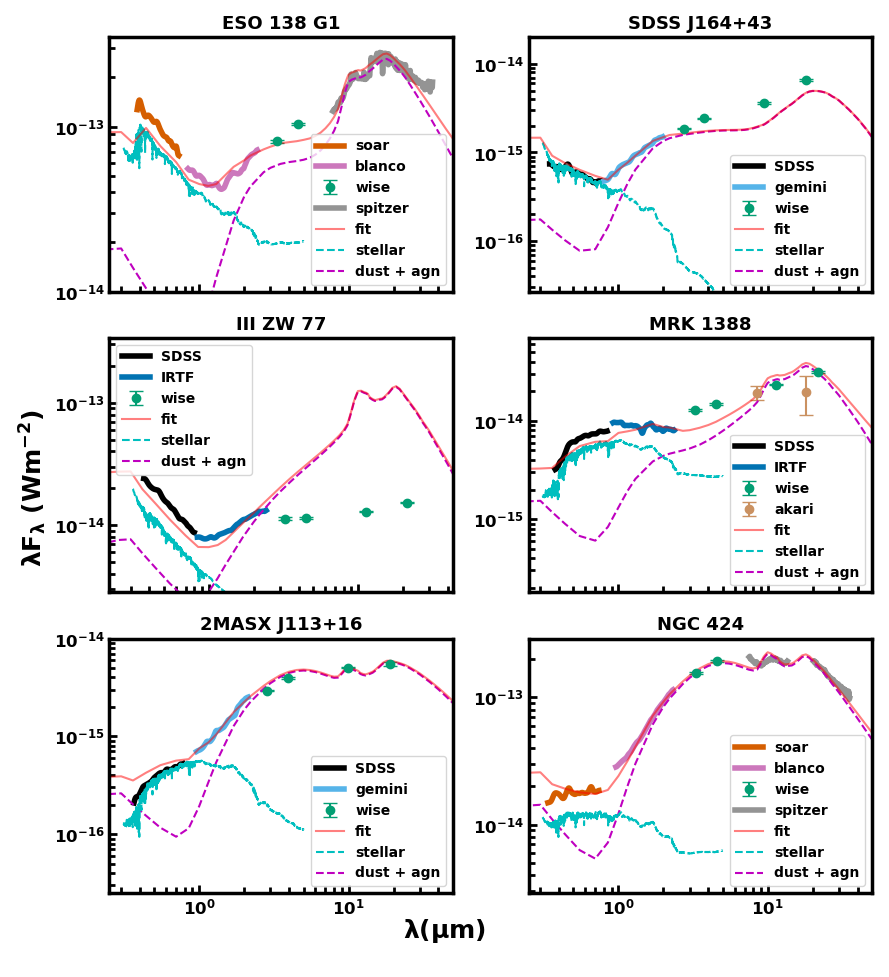}
    \caption{Best fit to CLiF AGN SEDs to {\sc cat3d-wind} models. The cyan line represents the stellar component, the dashed magenta line represents the dust plus the central source emission and the red solid line is the best fit. The x-axis represents the rest wavelength.}
	\label{fig:sed3}
	\centering
\end{figure*}

\begin{figure*}
    \includegraphics[width=\textwidth]{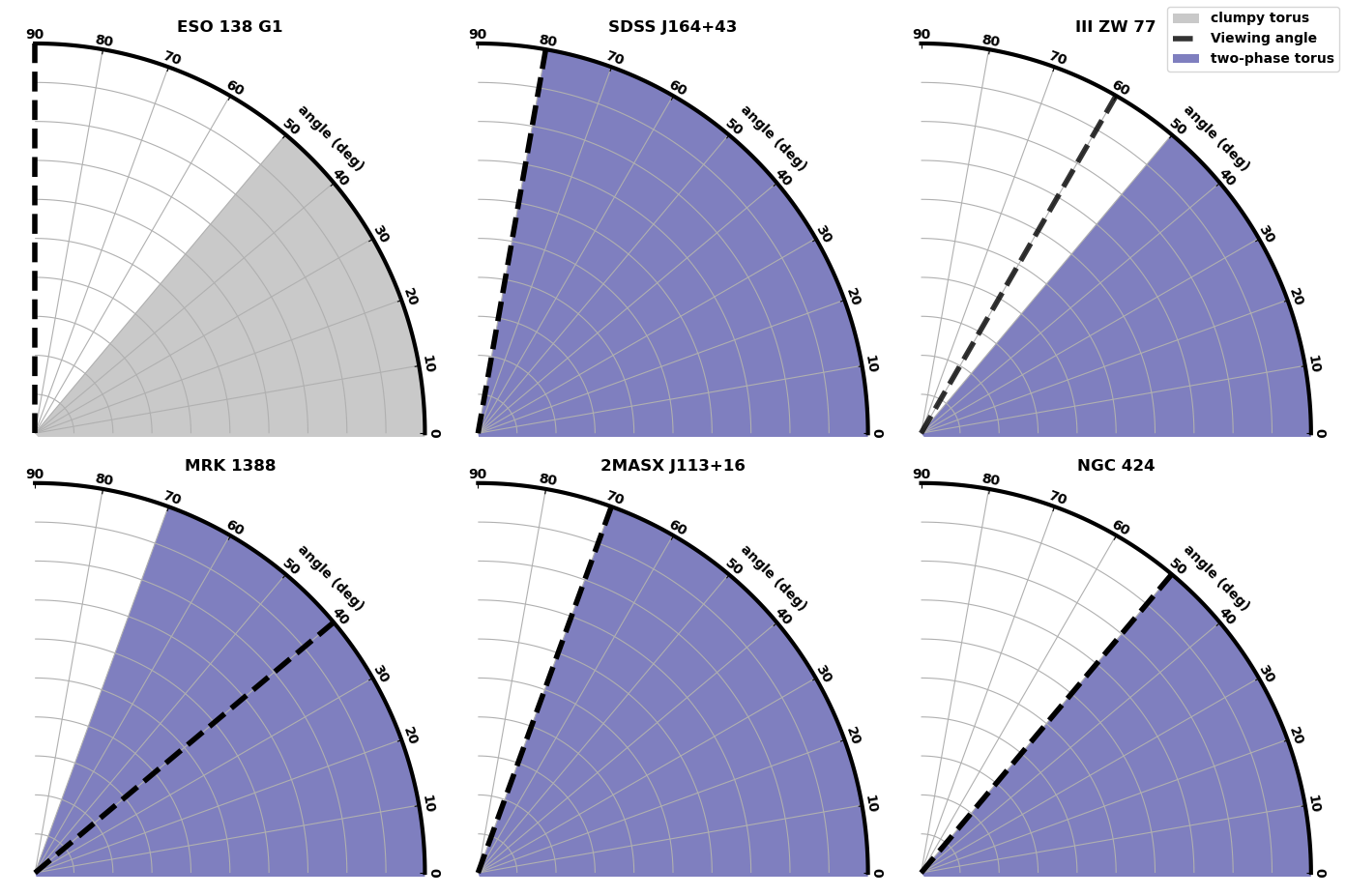}
    \caption{The representation of the parameters of the best fits. For fits made with {\sc clumpy} the grey area represents the half-opening angle of the clumpy torus ($\sigma$). For fits made with {\sc skirtor}, the half-opening angle of the two-phase torus ($\sigma$) is in purple. In all panels, the viewing angle ($i$) is represented by a black dashed line.}
	\label{fig:polartodos}
	\centering
\end{figure*}

Figures \ref{fig:sed1}, \ref{fig:sed2} and \ref{fig:sed3} show the result of the SED fitting to our galaxy sample. The observational data described in Section \ref{dados} is labelled in the figure. The dashed light-blue line represents the stellar component, the dashed magenta line represents the dust emission, the dashed dark-blue line is the scattered disk component, and the red solid line is the best fit. Tables \ref{tab:sedscat3d} and \ref{tab:sedsskir} list the values of the parameters found for the best fit for each of these two approaches. Figure~\ref{fig:polartodos} displays a planar diagram with the parameters $i$ and $\sigma$ found to the best-fit SED models for CLiF AGNs. For this figure fits made with {\sc clumpy} the grey area represents the parameter $\sigma$, fits made with {\sc skirtor}, the representation of $\sigma$ is in purple. In all panels of the Figure~\ref{fig:polartodos}, the parameter $i$ is represented by a black dashed line.

\begin{table*}
\centering
\caption{Values of the parameters for the galaxy which obtained the best fit with the {\sc clumpy} model} 
\begin{tabular}{lcccccc}
\hline
\multicolumn{1}{c}{Galaxy} & N$_{0}$    & Y           & q             & $\sigma$ (deg) & $\tau_{\nu}$     & $i$ (deg)      \\ \hline \\
ESO 138 G1                 & 14 $\pm$ 1 & 30 $\pm$ 10 & 0.5 $\pm$ 0.5 & 50 $\pm$ 5     & 20$_{-10}^{+20}$ & 0$_{-0}^{+10}$ \\ \\ \hline
\end{tabular}
\label{tab:sedscat3d}%
\end{table*}

\begin{table*}
\centering
\caption{Values of the parameters for each galaxy which obtained the best fit with the {\sc skirtor} model} 
\begin{tabular}{lcccccc}
\hline
Galaxy        & \multicolumn{1}{c}{$i$ (deg)}   & \multicolumn{1}{c}{$\tau _{9.7\mu m}$} & $\sigma$ (deg) & $p$                 & $q$                 & $Y$              \\ \hline \\
SDSS J164+43  & 10 $\pm$ 10                     & \textbf{9 $\pm$ 2}                              & 80 $\pm$ 10    & 1.0 $\pm$ 0.5       & 0.0$_{-0.0}^{+0.5}$ & \textbf{20$_{-10}^{+10}$} \\[6pt]
III ZW 77     & \textbf{30 $\pm$ 10}                     & 5 $\pm$ 2                              & 50 $\pm$ 10    & 1.0 $\pm$ 0.5       & 0.0$_{-0.0}^{+0.5}$ & \textbf{20$_{-10}^{+10}$ }\\[6pt]
MRK 1388      & 50 $\pm$ 10                     & 5 $\pm$ 2                              & 70 $\pm$ 10    & 0.0$_{-0.0}^{+0.5}$ & 1.5$_{-0.5}^{+0}$   & 30$_{-10}^{+0}$  \\[6pt]
2MASX J113+16 & \multicolumn{1}{c}{\textbf{20 $\pm$ 10}} & \multicolumn{1}{c}{7 $\pm$ 2}          & 70 $\pm$ 10    & 1.5$_{-0.5}^{+0}$   & \textbf{0.5 $\pm$ 0.5}       & 30$_{-10}^{+0}$  \\[6pt]
NGC 424       & \multicolumn{1}{c}{40 $\pm$ 10} & \multicolumn{1}{c}{\textbf{9 $\pm$ 2}}          & 50 $\pm$ 10    & 0.0$_{-0.0}^{+0.5}$ & \textbf{0.5 $\pm$ 0.5}       & 10$_{-0}^{+10}$  \\ \\ \hline
\end{tabular}
\label{tab:sedsskir}%
\end{table*}

A close inspection of the SED shapes in figures \ref{fig:sedst1e2skir} and \ref{fig:sedst1e2cat}  obtained for NGC\,1068 and NGC\,4151, respectively, with the best fitted SED shapes of the CLiF AGNs (blue panels in figures \ref{fig:sed1} and \ref{fig:sed2}) suggests that all but 2MASX\,J113+16 and NGC\,424 have a SED that is very close to that of NGC 4151, this latter object assumed as representative of Type~I AGN. It is important to notice that this latter SEDs is characterized by a flatter shape than that of a Type~II AGN. It also shows the presence of silicate emission except in SDSS\,J164+43, where it is in absorption as suggested by {\sc skirtor} model (Figure \ref{fig:sed2}).

Tables \ref{tab:sedscat3d} and \ref{tab:sedsskir} show that the torus inclination determined by {\sc clumpy} and {\sc skirtor} covers the range of values from 0$^{\circ}$ to 50$^{\circ}$. Overall, the infrared emission is compatible with that of a Type~I AGN in most objects of the sample. 

\section{Comparison of model parameters between CLiF and non-CLiF AGNs}\label{compAnelise}

\defcitealias{gonzalez/2019}{GM19}\citet[][hereafter GM19]{gonzalez/2019} carried out an infrared SED fitting of 110 AGNs using the Spitzer/IRS spectroscopic data with six different models, including {\sc skirtor} and {\sc clumpy}. As their sample is fully dominated by non-CLiF AGNs and contains a significant number of objects, it is worth comparing their results with ours. The distribution of the parameter space found for {\sc skirtor} and {\sc clumpy} in \citetalias{gonzalez/2019} are illustrated in their figures 15 and 18. The main results of the comparison between each parameter using the \citetalias{gonzalez/2019} as a reference, can be summarized as follows:

\begin{itemize}

\item The viewing angle, $i$: The five objects best fitted in this work with {\sc skirtor} display torus inclination angle $i$ between 10 and 50~deg. In the \citetalias{gonzalez/2019} distribution the least frequent values of $i$ are 20 and 30~deg, which are the values obtained for SDSS\,J164+43 and III\,ZW\,77, respectively. Moreover, the AGN that is best-fit with {\sc clumpy} (ESO\,138\,G1), has an $i$-value of 0~deg. It is located in the second largest peak of the distribution in \citetalias{gonzalez/2019}.

\item The number of dusty clouds in the direction of the line of sight, $N_{0}$: In the only AGN that was fit with {\sc clumpy} (ESO\,138\,G1) the value of  $N_{0}$=14 is close to the highest peak of the \citetalias{gonzalez/2019} distribution.

\item The ratio between the outer and inner radius of the torus,  ($Y$): The values of 10 and 30 obtained here with {\sc skirtor} coincide with the two peaks of the distribution in \citetalias{gonzalez/2019}. In SDSS\,J164+43 and III\,ZW\,77, the $Y$ of 10 and 30, respectively, is not present in their distribution. In ESO\,138\,G1 ({\sc clumpy}), the $Y$ value of 30 is present in a secondart peak, however not very frequent.

\item The radial density of clouds of the torus, $q$ ({\sc clumpy}) and $p$ ({\sc skirtor}): The value of 0.1 and 1.5 found using {\sc skirtor} coincides with the peaks of the distribution of \citetalias{gonzalez/2019}, with the value 1.5 being within the largest peak of the distribution. For the one AGN fit with {\sc clumpy}, the value of 0.5 is very frequent.

\item The half-opening angle, $\sigma$: The results from {\sc skirtor} provided values of 50, 70, and 80 deg, coinciding with peaks in the distribution in \citetalias{gonzalez/2019}. Moreover, the distribution of values of $\sigma$ obtained with {\sc clumpy} in \citetalias{gonzalez/2019} is quite homogeneous, with a peak at  70 deg. The value of 50 deg obtained for ESO\,138\,G1 does not correspond to any of the peaks in the \citetalias{gonzalez/2019} distribution.

\item The index of the dust density gradient with polar angle function, $q$: 
In half the sample, the values of $q$ derived from {\sc skirtor} coincide with peaks of the distribution shown in \citetalias{gonzalez/2019}. In the remaining objects $q$ are out of the peaks but still are within the range of possible values displayed by non-CLiF AGN.
 
\item The edge-on optical depth at 9.7 microns, $\tau _{9.7\mu m}$: The values of 5, 5 and 7 found for III\,ZW\,77, MRK\,1388 and 2MASX\,J113+16, respectively, coincide with the peaks of the distribution in \citetalias{gonzalez/2019}. However the value of 9 present in SDSS\,J164+43 and NGC\,424 is less frequent.

\item The optical depth of the individual clouds ($\tau$): For ESO\,138\,G1 ({\sc clumpy}), the value of 20 is frequent in the distribution of \citetalias{gonzalez/2019}.

\end{itemize}

In summary, for the set of parameters that characterize the SED fitting in {\sc clumpy} and {\sc skirtor}, the CLiF AGNs do not display values that stand out from the distribution of non-CLiF AGNs. With the exception of some parameters that differ from the distribution of non-clif AGNs (indicated in bold in Table \ref{tab:sedsskir}), but do not have a distinct relationship in the sample as a whole. It means that the objects of our sample do not represent a class with very different characteristics from other types of AGNs regarding the parameters of dusty torus models.

In \citetalias{cerqueira/2021}, it was found that the physical properties of the emission line gas in CLiF AGNs do not differ from those of non-CLiF AGNs, meaning that the former is not a separate group of AGN. Moreover, it was stated that a specific viewing angle does not define the presence of a strong coronal line forest. They also found that CLiF AGNs are dominated by Type~I objects and that the black hole mass of the AGN is $<10^8$~M$_{\odot}$. Finally, the study of their kinematics suggests a compact NLR. In this work, by means of the SED fitting in the optical/infrared, we confirm that in CLiF AGN, the line of sight of the observer is directed towards the central engine (see Figure \ref{fig:polartodos}).

It is possible to notice, by looking at Figure \ref{fig:polartodos}, that Mrk\,1388 is the only object fit with {\sc skirtor} where the line of sight to the central engine is intercepted by the bulk of the torus. This result contrasts to what was verified in \citetalias{cerqueira/2021}. It displays a broad component (FWHM of 1040 km~s$^{-1}$) in the Balmer lines, while \cite{doi/2015} observed a strong featureless continuum in the optical spectrum. This would tentatively classify it as a narrow line Seyfert~1 (NLS1). However, a similar broad component is also detected (in blue-shift) in the \OIII\ lines, which suggests the presence of an outflow rather than a genuine BLR emission line. In the NIR, broad components were not identified in the lines of \ion{H}{i}. The results obtained in this work suggest that Mrk\,1388 is a Type~II AGN. 

For ESO\,138\,G1, the viewing angle points out that we are looking directly to the central engine. However, the number of clouds in the line of sight is high (N$_{0}$ = 14 $\pm$ 1). This would heavily obscure the central source, explaining its Type~II-type appearance. In SDSS\,J164+43, 2MASX\,J113+16, and NGC\,424 the results suggest a viewing angle that is grazing the torus,  allowing the view of the central source. Likewise, in III\,Zw\,77 the viewing angle allows a direct view  to the central source, in agreement with the detection of a truly BLR. These results indicate that although we have a direct view to the central region, the inclination angle points to regions influenced by dust. This then allows a scenario of dust grains being destroyed, favoring the observation of a greater number of coronal lines. It is also in line with the fact that some CLiF AGN has a hidden BLR in the optical region (\citetalias{cerqueira/2021}). We are aware of the small number of of sources currently identified as CLiF AGN and this may affect these results somehow. 

To draw more robust conclusions about the nature of the NLR in objects with coronal line forests, it is necessary to identify more CLiF AGN and expand the current analysis to them. The analysis carried out so far in this work as well as in \citetalias{cerqueira/2021} shows that the presence of a coronal line forest is not related to either a special viewing angle or a very exotic property of the emission gas.

\section{Final remarks}\label{conclusions}

In this work, we carry out optical plus infrared SED fitting in a sample of 6 galaxies classified as CLiF AGN using three different torus models, namely {\sc clumpy}, {\sc skirtor}, and {\sc cat-3d}. This approach allowed us to determine the orientation angle of the torus with respect to the observer for all objects. The results show that the viewing angle  is compatible with the orientation of a Type~I AGN, with the exception of one source (MRK\,1388). We also found from the SED fitting the presence of dusty clouds in the line of sight. For the source best fitted with {\sc clumpy} (ESO\,138\,G1), we found the presence of dust clouds in the polar region. We also found that in three objects of the sample (SDSS\,J164+4, 2MASX\,J113+16, and NGC\,424) the fits made with {\sc skirtor} show that the viewing angle is such that it coincides with the half-opening angle of the torus. This would explain the presence of broad components in the permitted lines, likely associate to the BLR and an spectrum that overall looks like a Seyfert~2 in the optical.

Comparing the results obtained using {\sc clumpy} and {\sc skirtor} for CLiF and non-CLiF AGNs, we found that CLiF and non-CLiF AGNs behave similarly. This result shows that CLiFs cannot be treated as a separate class of  AGNs. However, the number and intensity  of coronal lines detected in the optical and NIR make CLiFs critical targets for understanding the processes behind the production of coronal lines and the physical conditions of AGNs in general \citep[see][for a recent study]{prieto/2022}.

The number of CLiF AGNs classified as such up to today is small (7 in total). More objects with similar properties need to be identified in order to draw definitive conclusions about the nature of the coronal line region in CLiF AGN. The study of the known sample of such objects, in most cases, points out that they are compatible with Type~I AGN orientation with significant influence of dusty clouds, favouring CL emission.

\section*{Acknowledgements}

We thank the anonymous referee for their useful comments and suggestions that helped to improve this manuscript. 
FCCC acknowledges the PhD grant from CAPES. RR acknowledges support from the Fundaci\'on Jes\'us Serra and the Instituto de Astrof{\'{i}}sica de Canarias under the Visiting Researcher Programme 2023-2025 agreed between both institutions. RR, also acknowledges support from the ACIISI, Consejer{\'{i}}a de Econom{\'{i}}a, Conocimiento y Empleo del Gobierno de Canarias and the European Regional Development Fund (ERDF) under grant with reference ProID2021010079, and the support through the RAVET project by the grant PID2019-107427GB-C32 from the Spanish Ministry of Science, Innovation and Universities MCIU. This work has also been supported through the IAC project TRACES, which is partially supported through the state budget and the regional budget of the Consejer{\'{i}}a de Econom{\'{i}}a, Industria, Comercio y Conocimiento of the Canary Islands Autonomous Community. RR also thanks to Conselho Nacional de Desenvolvimento Cient\'{i}fico e Tecnol\'ogico  ( CNPq, Proj. 311223/2020-6,  304927/2017-1 and  400352/2016-8), Funda\c{c}\~ao de amparo \`{a} pesquisa do Rio Grande do Sul (FAPERGS, Proj. 16/2551-0000251-7 and 19/1750-2), Coordena\c{c}\~ao de Aperfei\c{c}oamento de Pessoal de N\'{i}vel Superior (CAPES, Proj. 0001). ARA acknowledges CNPq for partial support of this project. SP acknowledge the partial financial support from the Conselho Nacional de Desenvolvimento Científico e Tecnológico (CNPq) Fellowship (164753/2020-6) and the Polish Funding Agency National Science Centre, project 2017/26/A/ST9/-00756 (MAESTRO 9). Based on observations obtained at the Gemini Observatory, which is operated by the Association of Universities for Research in Astronomy, Inc., under a cooperative agreement with the NSF on behalf of the Gemini partnership: the National Science Foundation (United States), National Research Council (Canada), CONICYT (Chile), Ministerio de Ciencia, Tecnolog\'{i}a e Innovaci\'{o}n Productiva (Argentina), Minist\'{e}rio da Ci\^{e}ncia, Tecnologia, Inova\c{c}\~{o}es e Comunica\c{c}\~{o}es (Brazil), and Korea Astronomy and Space Science Institute (Republic of Korea). This paper is also based on observations obtained at the Southern Astrophysical Research (SOAR) telescope, which is a joint project of the Minist\'{e}rio da Ci\^{e}ncia, Tecnologia, Inova\c{c}\~{o}es e Comunica\c{c}\~{o}es (MCTIC) do Brasil, the U.S. National Optical Astronomy Observatory (NOAO), the University of North Carolina at Chapel Hill (UNC), and Michigan State University (MSU).

\section*{DATA AVAILABILITY}

Data products in the wavelength interval 1-2.5 microns will be shared on reasonable request to the corresponding author. Data employed in this work in the IR were downloaded from public databases of these facilities. The raw data products of the Spitzer spectra are available on the public Spitzer Heritage Archive at \url{https://sha.ipac.caltech.edu/applications/Spitzer/SHA/}. The  photometric points from the space telescopes WISE and AKARI are available on the VizieR photometry tool at \url{http://vizier.unistra.fr/vizier/sed/}. Processed data products underlying this article will be shared on reasonable solicitation to the authors.




\bibliographystyle{mnras}
\bibliography{mnras_template} 





\bsp	
\label{lastpage}
\end{document}